\documentclass[aps,prd,final,twocolumn,letterpaper, superscriptaddress]{revtex4}

\usepackage{amsmath, amssymb, amsthm, cancel, comment, dsfont, enumerate, epstopdf, eucal, fancyhdr, feynmf, gensymb, graphicx, lipsum, mathtools, physics, tikz, xcolor,verbatim}

\begin{document}
\title{High-Energy Astrophysical Neutrinos from Cosmic Strings}
\author{Cyril Creque-Sarbinowski}
\email{creque@jhu.edu}
\affiliation{William H.\ Miller III Department of Physics and Astronomy, Johns Hopkins University, 3400 N. Charles St., Baltimore, Maryland 21218, USA}
\author{Jeffrey Hyde}
\affiliation{Department of Physics and Astronomy, Bowdoin College, 8800 College Station, Brunswick, Maine 04011-8488, USA}
\affiliation{Department of Physics, Moravian University, 1200 Main Street, Bethlehem Pennsylvania 18018, USA}
\author{Marc Kamionkowski}
\affiliation{William H.\ Miller III Department of Physics and Astronomy, Johns Hopkins University, 3400 N. Charles St., Baltimore, Maryland 21218, USA}
\date{\today} 

\begin{abstract}
Cosmic strings that couple to neutrinos may account for a portion of the high-energy astrophysical neutrino (HEAN) flux seen by IceCube. Here, we calculate the observed spectrum of neutrinos emitted from a population of cosmic string loops that contain quasi-cusps, -kinks, or kink-kink collisions. We consider two broad neutrino emission models: one where these string features emit a neutrino directly, and one where they emit a scalar particle which then eventually decays to a neutrino. In either case, the spectrum of cosmic string neutrinos does not match that of the observed HEAN spectrum. We thus find that the maximum contribution of cosmic string neutrinos, through these two scenarios, to be at most $\sim 45$\% of the observed flux. However, we also find that the presence of cosmic string neutrinos can lead to bumps in the observed neutrino spectrum. Finally, for each of the models presented, we present the viable parameter space for neutrino emission.  
\end{abstract}

\maketitle

\pagestyle{myheadings}
\markboth{Cyril}{High-Energy Astrophysical Neutrinos from Cosmic Strings}
\thispagestyle{empty}

\section{Introduction}\label{sec:intro}

IceCube routinely detects high-energy astrophysical neutrinos (HEANs) with TeV-PeV energies that follow a power law flux spectrum with spectral index $\gamma = 2.53$~\cite{1907.11266}. Explanations for the source of this flux have ranged from gamma-ray bursts~\cite{astro-ph/9701231, 0907.2227, 1101.1448, 1204.4219, 1412.6510, 1601.06484, 1702.06868}, FR0 quasars~\cite{1711.03757}, blazars~\cite{1611.03874, 1810.02823, 1904.06371}, radio-bright AGN~\cite{2001.00930, 2009.08914, 2103.12813}, choked jet supernovae~\cite{1706.02175, 1809.09610}, pulsar wind nebulae~\cite{2003.12071}, and more. However, none of these propositions have been succesful at explaining the majority of the observed spectrum~\cite{2008.04323}. One additional possibility is that cosmic string loops source these neutrinos. More concretely, the actual mechanism of emission could be due to the radiation of particles from string features, known as quasi-cusps, -kinks, or kink-kink collisions, that generically occur during the evolution of cosmic string loops. These particles could either be the neutrinos themselves (direct neutrino emission) or a parent particle which then decays into neutrinos (indirect neutrino emission). 

The emission of neutrinos due to the decay of a real scalar radiated from cusps and kinks has previously been considered in the ultra-high energy range~\cite{1108.2509, 1206.2924}. Moreover, the energy spectrum of various Standard Model (SM) particles near the string has been extensively computed in the context of dark strings coupling through Higgs portal operator~\cite{1312.4573, 1405.7679, 1409.6979}. More generally, the program of calculating emission from cosmic strings also includes the radiation of gravitational waves, cosmic rays, and more~\cite{Vachaspati:1984gt, Hindmarsh:1990xi, Allen:1991bk, 0911.2655, gr-qc/0104026, 1911.12066}.  

In this work we extend and refine these calculations in several manners. First, we calculate the optical depth of HEANs using all seven channels of Standard Model neutrino self-interactions and thus including the energy dependence of the neutrino horizon. Then, we perform this calculation for all three types of string features: quasi-cusps, -kinks, and kink-kink collisions. Prior work has only considered the first two in the scenario of neutrino emission. In addition, we calculate the emission from a real scalar not only in the scenario of a cascade of particles, but also the direct decay into neutrinos. Moreover, we present the first calculation for the emission of neutrinos directly from cosmic strings via a 2-body decay interaction and the Aharonov-Bohm coupling. Finally, we incorporate the shrinking of loops due to particle radiation into the loop distribution function, a factor ignored in earlier neutrino emission papers. Using these calculations, we present the viable parameter space for neutrino emission for each of the models chosen. Moreover, we find that for these models, cosmic strings can only contribute at most around $45\%$ of the observed HEAN flux. Since the models represent a wide selection of possible emission mechanisms, we conclude it is unlikely that a single population of cosmic strings can create the entirety of the HEAN background. However, as a subdominant component, cosmic strings may still contribute enough to create a bump in the spectrum.  

This paper is organized as follows. In Sec.~\ref{sec:nu_flux_form} we present the general formalism in order to calculate the differential flux of neutrinos observed at IceCube from an arbitrary source and then particularize to the case of a cosmic string loop population. For this population, we introduce four interaction terms between cosmic strings and neutrinos using an effective field theory approach in Sec.~\ref{sec:cs_pheno}. These interactions cover both direct and indirect neutrino emission, each of which is split into two further cases. We use these interaction terms to then calculate the energy spectrum of neutrinos emitted at the locality of the string in Sec.~\ref{sec:emission}. We follow up this calculation and then specify the form of the cosmic string loop number density in Sec.~\ref{sec:string_pop}. Ultimately, we combine both the energy spectrum of neutrinos with the cosmic string loop number density to calculate the observed differential flux of neutrinos through the formalism presented in the beginning, shown in Eq.~\eqref{eq:maineq}. Using this flux, we constrain both the fraction of neutrinos attributed to emission from cosmic strings in the IceCube spectrum and the phenomenological parameter space for neutrino emission in Sec.~\ref{sec:results}. We discuss and conclude in Sec.~\ref{sec:disc} and \ref{sec:conc}. 
\section{Neutrino Specific Flux}\label{sec:nu_flux_form}
The specific flux  $\Phi_i(t, E)$ of neutrinos $\nu_i$ (number of astrophysical neutrinos per unit conformal time per unit comoving area per unit energy) at cosmic time $t$ and observed energy $E$ from a source $S_i$ is \cite{2005.05332}
\begin{align}\label{eq:boltzmann}
\Phi_i(t, E) = \int_{-\infty}^t dt'[a(t)/a(t')]e^{-\tau_i(t', t, E)}S_i\{t', [a(t)/a(t')E\},
\end{align} 
where $a(t)$ is the scale factor and $\tau_i(t', t, E)$ is the optical depth of a neutrino $\nu_i$ of energy $E$ between times $t'$ and $t$. 

\begin{figure}
\includegraphics[width = 0.50\textwidth]{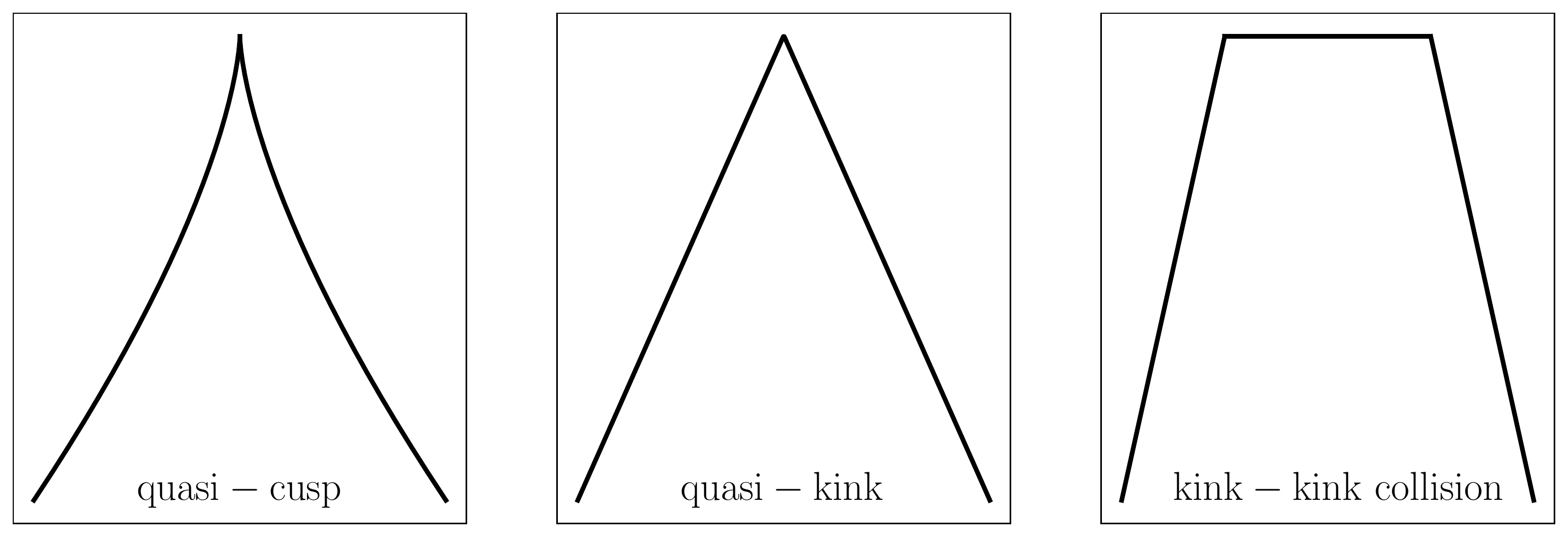}
\caption{Picture of quasi-cusps, quasi-kinks, and kink-kink collisions. }\label{fig:features}
\end{figure}

For a single cosmic string loop, the spectrum of emitted neutrinos is a function of the loop length $L$, and so the source function is the integral over all loop contributions,
\begin{align}\label{eq:gen_source}
S_{i, a}^{e}(t, E) &= \sum_a c\int_0^{\infty} dL\frac{d\dot{N}^{e}_{i, a}(t, L, E)}{dN_{\rm loop} dE}\frac{dn_{\rm loop}(t, L)}{dL},
\end{align}
with $dn_{\rm loop}(t, L)/dL$ the number of cosmic string loops per comoving volume per loop length, and $d\dot{N}^{e}_{i, a}/dN_{\rm loop}dE$ the number of neutrinos ultimately produced from string feature $a$ and emission model $e$ per unit time per loop per neutrino energy $E$. The string features we consider are quasi-cusps, quasi-kinks, and kink-kink collisions, shown in Fig.~\ref{fig:features}, so that the label $a$ takes values $a \in \{qc, qk, kk\}$. We present the different emission models in Sec.~\ref{sec:cs_pheno}. In general, a loop can contain multiple features at once (e.g. a string could have 4 quasi-kinks and and quasi-cusp). Here, for simplicity, we assume that only a single feature exists on every loop.  We then write the emitted neutrino spectrum as
\begin{align}
\frac{d\dot{N}_{i, a}^e(t, L, E)}{dN_{\rm loop} dE} &= \frac{1}{[(L/2)/c]}\int dE_p \frac{dN_i^e(E, E_p)}{dN_a^e dE}\frac{dN_a^e(E_p , L)}{dE_p},
\end{align} 
with $[(L/2)/c]$ the period of oscillation for a cosmic string loop, $dN_{i}^e/dN_a^edE$ the number of neutrinos emitted per parent particle per unit neutrino energy $E$, and $dN_a^e/dE_p$ the number of parent particles emitted from string feature $a$ per unit parent particle energy $E_p$. 

If neutrinos are emitted directly from the cosmic string and there is no parent particle, we set $dN_i^e(E, E_p)/dN_a^edE(E, E_p) = \delta(E - E_p)\delta_e^i$ with $\delta(x)$ the Dirac delta function and $\delta_i^j$ the kronecker delta function that determines if the neutrino $i$ is the same as the emitted particle in emission model $e$.   

Roughly speaking, the cosmic string phenomenology is then encoded in the emitted neutrino spectrum, and the cosmic string population dynamics in its number density.

\begin{figure}
\includegraphics[width = 0.50\textwidth]{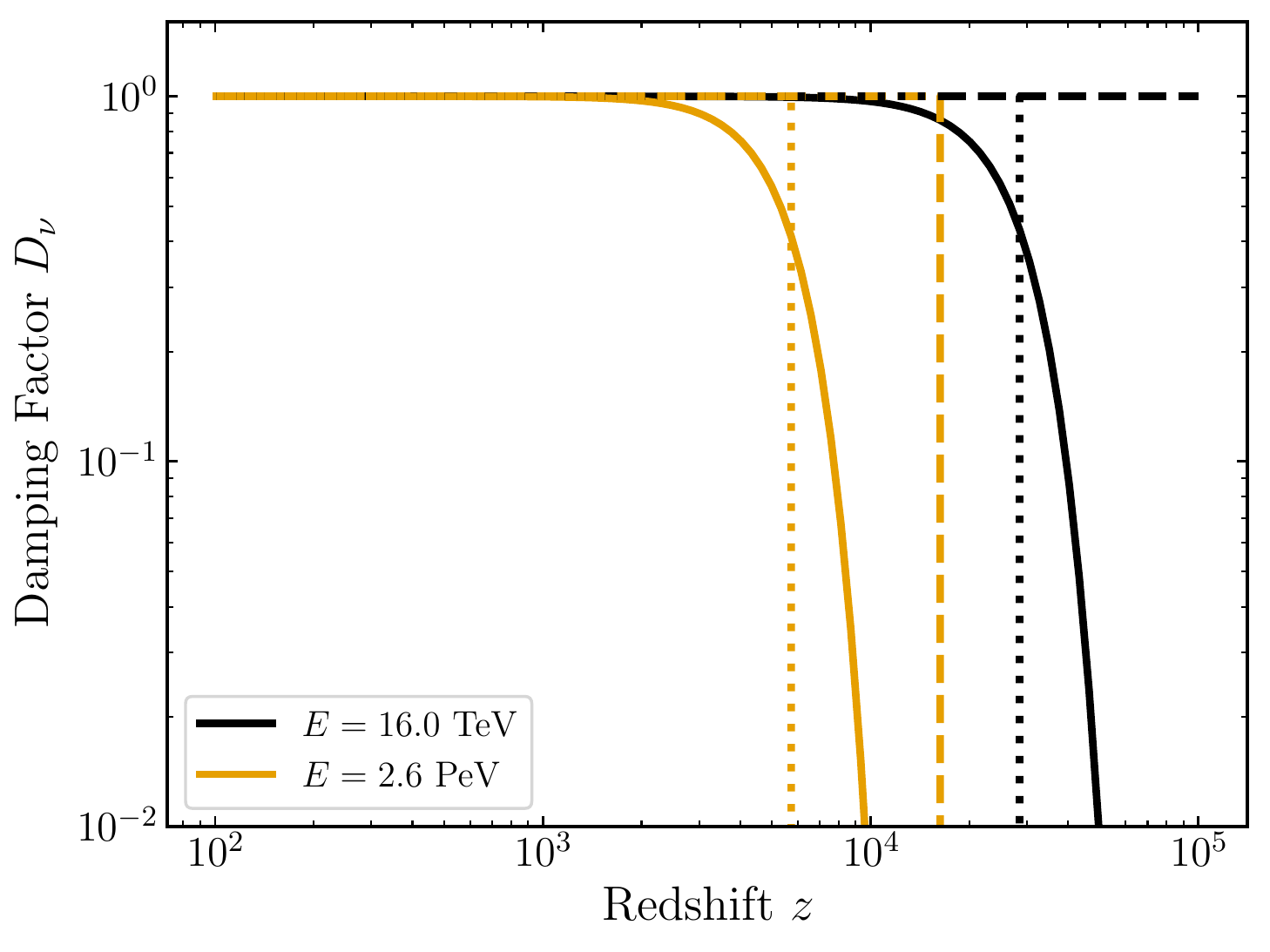}
\caption{The HEAN damping factor $D_{\nu_i}$. The solid lines indicate the complete damping factor, the dashed the approximation given by Ref.~\cite{1108.2509}, and the dotted are given by our approximation in Eq.~\eqref{eq:Dnu}}\label{fig:Dnu}
\end{figure}

Neutrino self-interactions ($\nu$SI) in the Standard Model (SM) induce scattering between HEANs and cosmic background neutrinos and thus a nonzero HEAN optical depth. We evaluate the total HEAN optical depth following Ref.~\cite{1312.3501}, including all seven channels of SM $\nu$SI. These channels lead to a sharply defined neutrino horizon at redshift $z_{\nu_i}$. That is, an observer located at redshift $z(t)$ will not see neutrinos of a given energy $E$ originating from a redshift $z(t') > z_{\nu, i}(t, E)$. Therefore, in order to simplify our expressions, we will take the following approximation
\begin{align}\label{eq:Dnu}
D_{\nu_i}(t', t, E) \equiv e^{-\tau_i(t', t, E)} = \Theta[z_{\nu_i}(t, E) - z(t')],
\end{align} 
for the damping factor, with $z_{\nu_i}$ defined by the expression $D_{\nu_i}\{t'[z_{\nu, i}(t, E)], t, E\} = \exp(-1)$. We show both the complete HEAN optical depth and our approximation in Fig.~\ref{fig:Dnu} for some typical energies. 

\section{Cosmic String Phenomenology}\label{sec:cs_pheno}
Cosmic strings are topological defects formed after a $U(1)$ symmetry-breaking phase transition occurs in the Universe and are characterized by their tension $\mu$. Then there are two broad phenomenological categories by which neutrinos may be emitted from this string. First, the string may directly couple to neutrinos. Second, it may indirectly couple to neutrinos; i.e. it may emit some intermediary particle which then eventually converts to some number of neutrinos. In either case, we model the free string loop action using the Nambu-Goto action for a infinitely long straight string, as locally the string loop is straight, regardless of any features 
\begin{align}
\mathcal{L}_{\rm str} &= -\frac{\mu}{\hbar c}\int d^2\sigma \sqrt{-\gamma} \delta^4[x^\mu - X^\mu(\sigma, \tau)],
\end{align} 
where $\hbar$ is Planck's constant, $g \equiv \det g_{\mu\nu}$ is the determinant of the spacetime metric $g_{\mu\nu}$, and $\gamma$ the analogous quantity for the induced worldsheet metric $\gamma_{ab} = g_{\mu\nu}X^\mu_{, a}X^\nu_{,b}$ with worldsheet coordinates $(\sigma, \tau)$. We take the background metric to be flat $g_{\mu\nu} = \eta_{\mu\nu} = {\rm diag}(-1 ,1, 1, 1)$. 

This string then has stress-energy tensor 
\begin{align}
T^{\rm str}_{\mu\nu}(x^\mu)&= \frac{-\mu}{\sqrt{- g}}\int d^2\sigma \sqrt{-\gamma}\gamma_{ab}X^{,a}_\mu X^{,b}_\nu\delta^4[x^\mu - X^\mu(\sigma, \tau)]
\end{align}
with trace
\begin{align}
T_{\rm str}(x^\mu)&= -2\mu\int d^2\sigma\sqrt{-\gamma}\delta^4\left[x^\mu - X^\mu(\sigma, \tau)\right],
\end{align}
where we neglect any backreaction of interactions onto the string as for the models we consider they are small. When considering interactions with this string we take an effective field theory approach and remain agnostic to any particular ultraviolet theory constraints.
\subsection{Direct Coupling}\label{subsec:direct_pheno}
 For simplicity, we consider only a single neutrino species of mass $m_\nu = \hbar/(c\lambda_\nu)$ and take it to be a Dirac fermion. Thus its free Lagrangian is 
\begin{align}
\mathcal{L}_{\rm free}^\nu &= -\bar{\nu}\left(i\gamma^\mu\partial_\mu - \lambda_\nu^{-1}\right)\nu,
\end{align}
with $\gamma^\mu$ the gamma matrices. There are two versions of direct cosmic string coupling we consider. First, the neutrinos may couple directly to the string worldsheet through a two-body interaction, so that its interaction is
\begin{align}
\mathcal{L}_{\rm int}^{(2)} &= \frac{g^{(2)}}{2}\left(\frac{\hbar c}{\mu^3}\right)^{1/2} \bar{\nu}\nu T_{\rm str},
\end{align} 
with $g^{(2)}$ the two-body interaction coupling. Second, neutrinos may couple through some gauge flux that permeates through the string in an Aharonov-Bohm (AB) fashion \cite{0911.0682}
\begin{align}
\mathcal{L}_{\rm int}^{\rm AB} &= g_\nu \bar{\nu}\gamma^\mu V_\mu\nu,
\end{align}
with $g_\nu$ the charge of the neutrino under $V_\mu$, $V_\mu$ a classical background field induced by the flux $\Phi = (2\pi/g_\nu)\theta_q$ the string carries, and $\theta_q$ the AB phase around the string. In the Lorentz gauge, this background field is writen as \cite{Alford:1988sj}
\begin{align}
V_\mu &= -\frac{i\Phi}{2}\int \frac{d^4 k}{(2\pi)^4}\frac{p^\nu}{p^2}\int d\sigma_{\mu\nu}e^{-ik\cdot[x^\mu - X^\mu(\sigma, \tau)]},
\end{align} 
with $d\sigma_{\mu\nu} = d^2\sigma \epsilon^{\mu\nu\alpha\beta}\epsilon^{ab}X_{,a}^\alpha X_{,b}^\beta$ the worldsheet area element and $\epsilon^{i..j}$ the Levi-Civita symbol. Note that this field has support outside of the string, unlike the purely local interaction considered above. 

\subsection{Indirect Coupling}\label{subsec:indirect_pheno}
For indirect emission of neutrinos, we consider the intermediary particle to be a real scalar $\phi$ of mass $m_\phi = \hbar/(c\lambda_\phi)$. As a result, there is only one cosmic string Lagrangian to write down
\begin{align}
\mathcal{L} &= \mathcal{L}_{\rm str} + \mathcal{L}_{\rm free}^\phi + \mathcal{L}^\phi_{\rm int},\\
\mathcal{L}_{\rm free}^\phi &= -\left(\frac{1}{2}\partial_\mu \phi\partial^\mu\phi + \frac{1}{2}\lambda_\phi^{-2}\phi^2\right),\\
\mathcal{L}_{\rm int}^\phi &= \frac{\alpha}{(4 \mu \hbar c)^{1/2}}\phi T_{\rm str},
\end{align}
with, $\alpha$ is the scalar coupling constant.

In order to obtain neutrinos indirectly we consider two scenarios. First, the scalar particle decays directly into neutrinos via a Yukawa interaction
\begin{align}
\mathcal{L}_{\rm Yu} &= g_{\rm Yu} \bar{\nu}\phi\nu.
\end{align}
Alternatively, the scalar particle couples to some gauge boson - either a hidden sector gauge boson or the gluon, and these gauge fields have interactions which lead to a cascade of particles being emitted which end in neutrinos. For example, if it is the gluon, hadronic cascades produce pions which then lead to neutrino emission. For concreteness, we write down an example Lagrangian as
\begin{align}
\mathcal{L}_{\rm casc} &= \alpha \ell_{\rm P} \phi G_{\mu\nu}G^{\mu\nu}, 
\end{align}
with $\ell_{\rm p}$ the Planck length and $G_{\mu\nu}$ the gluon field strength tensor. 
\section{Particle Emission}\label{sec:emission}
Given a model for cosmic string interactions with neutrinos, we now write the number spectrum of particles emitted from cosmic string loops. However, this spectrum depends not only on the phenomenology of the interactions, but also the feature of the string that emits the particle. Thus, in what follows, for each interaction considered we specify the type of feature as well. 

In order to calculate the spectrum of emitted particles we take the leading-order $S$-matrix approach. Thus, we calculate the probability of creating a state $\bra{k_1, s_1; \ldots; k_N, s_N}$ with $N$ particles with momenta $k_i$ and spin $s_i$ out of the vacuum $\ket{0}$ given an interacting term, 
\begin{align}
\mathcal{A}_e({\bf k}, {\bf s})&= i\int d^4x \bra{ k_1, s_1;\ldots;k_N, s_N}\mathcal{L}^e_{\rm int}\ket{0}\\
dN_a^e &= \sum_{i = 1}^{N_s}\sum_{s_{a_i}}\prod_{j = 1}^N\frac{d^3k_j}{(2\pi)^2\omega_j}|\mathcal{A}_e({\bf k}, {\bf s})|^2,
\end{align}
with $N_s$ the number of particles with non-zero spins, $N$ the number of particles, $a_i$ the map from spin particle number to particle number (e.g. a particle could be the 1st particle with spin but the 5th overall particle in a list) and the sum $s_{a_i}$ goes over the possible spin values of particle $a_i$. Lower bounds on the energy of the resulting spectrum arise from integrating over the worldsheet. Upper bounds on the energy of the spectrum arise from the requirement that the energy of the particle is smaller than the string energy scale. For more details we refer the reader to Ref.~\cite{1405.7679}. While both of these cutoffs in reality have a slight softening, they still decay rapidly and so here we approximate them as sharp discontinuous transitions.  

The average power emitted from a cosmic string over one period of oscillation is therefore
\begin{align}\label{eq:power}
dP_a^e &= \frac{1}{[(L/2)/c]}\sum_{i = 1}^{N_s}\sum_{s_{a_i}}\prod_{j = 1}^N\frac{d^3k_j}{(2\pi)^2\omega_j}\left(\sum_{k = 1}^N \omega_k\right)|\mathcal{A}_e({\bf k}, {\bf s})|^2.
\end{align}
In order to complete the description of the string feature, several quantities must also be defined detailing the shape of the string feature in question. Rather than defining these quantities precisely, here we simply tabulate the numerical constants that encode their behavior, assuming that shape effects are $\mathcal{O}(1)$. Following this procedure, these constants then take a range of values: $\Theta \in [0.42, 3.6]$ and $\psi \in [0.047, 0.23]$. We define the rest of these constants in Table~\ref{table:shape}. For a first-principle definition of these parameters and their values we refer the reader to Ref.~\cite{1405.7679}.
\begin{table}[h!]
\begin{tabular}{|c|c|c|c|}
\hline
$a$ & $qc$ & $qk$ & $kk$\\
\hline
$\mathcal{S}_a$ & $[0.2, 10]$ & $[0.1, 20]$ & $[1, 500]$\\
$\mathcal{T}_a$ & $[0.5, 50]$ & $[1, 200]$ & $[0.2, 200]$\\
\hline
\end{tabular}
\caption{Range of values for cosmic-string shape-dependent variables, assuming the shape parameters are $\mathcal{O}(1)$.}\label{table:shape}
\end{table}
 
\subsection{Direct Coupling}
First, we present the spectrum of neutrinos directly emitted from cosmic string loops with quasi-cusps, quasi-kinks, and kink-kink collisions. 
\subsubsection{Two-Body}
For both quasi-cusps and quasi-kinks, the momenta of both emitted (nearly massless) fermions are parallel to one another, and thus the emission is helicity suppressed. For kink-kink collisions that emit relativistic neutrinos, 
\begin{align}
\frac{dN^{(2)}_{kk}}{dE} &= \tilde{\Gamma}_{kk}^{(2)}\frac{E}{\mu \hbar c}\left[1 + \left(\frac{E^2}{\mu \hbar c}\right)^{1/2}\right]^{-3},\\
P^{(2)}_{kk} &= \Gamma_{kk}^{(2)} \frac{\mu c}{L/\ell_{(2)}},
\end{align}
with $m_\nu c^2 \ll E \leq (\mu \hbar c)^{1/2}$ and $\tilde{\Gamma}^{(2)}_{kk} = 4 \left[g^{(2)}\right]^2 \mathcal{S}_{kk}/(3\pi^2)$, $\Gamma^{(2)}_{kk} = (37/5)\tilde{\Gamma}^{(2)}_{kk}$ and $\ell_{(2)} = (\hbar c/\mu)^{1/2} $.

\subsubsection{Aharonov-Bohm}
In AB emission, there are no obvious suppressions, and so we write down the spectrum and power for all  emission types in the relativistic limit,
\begin{align}
\frac{dN_a^{\rm AB}}{dE} &= \tilde{\Gamma}^{\rm AB}_{a}\left(\frac{\hbar c}{L}\right)^{q_a^{\rm AB}}\\
&\nonumber\times\left[\frac{1}{(E + E_{\rm min}^{{\rm AB}, a})^{1 + q_a^{\rm AB}}} - \frac{1}{(E + E_{\rm max}^{{\rm AB}, a})^{1 + q_a^{\rm AB}}}\right],\\
P_a^{\rm AB} &= \Gamma^{\rm AB}_{a}\frac{\mu c}{(L/\ell_{\rm AB})^{p_a^{\rm AB}}},
\end{align}
with $\tilde{\Gamma}_{qc}^{\rm AB} = (2\pi \theta_q)^2\psi^{-4/3}\Theta^2/[32(2\pi)^4]\mathcal{T}_{qc}$, $\tilde{\Gamma}^{\rm AB}_{qk} = [3\mathcal{T}_{qk}/(4\mathcal{T}_{qc})](2/\Theta)\tilde{\Gamma}^{\rm AB}_{qc}$, $\tilde{\Gamma}^{\rm AB}_{kk} = (\mathcal{T}_{kk}/\mathcal{T}_{qc})(2\Theta^2)^{-1}\psi^{4/3}\tilde{\Gamma}^{\rm AB}_{qc}$, and $\ell_{\rm AB} = (\hbar c/\mu)^{1/2}$. We define all other variables in Table~\ref{table:AB_vars}.
\begin{table}[h!]
\begin{tabular}{|c|c|c|c|}
\hline
$a$ & $qc$ & $qk$ & $kk$\\
\hline
$q_a^{\rm AB}$ & $0$ & $1/3$ & $0$\\
$p_a^{\rm AB}$ & $1/2$ & $4/3$  & $1$\\
$\Gamma^{\rm AB}_a$ & $\log(16)\tilde{\Gamma}^{\rm AB}_{qc}$ & $18(1 - 2^{-1/3})\tilde{\Gamma}^{\rm AB}_{qk}$  & $\log(16)\tilde{\Gamma}_{kk}^{\rm AB}$\\
$E_{\rm min}^{{\rm AB}, a}$ & $\psi m_\nu c^2\sqrt{m_\nu cL/\hbar}$ & $\psi m_\nu c^2\sqrt{m_\nu cL/\hbar}$  & $m_\nu c^2$\\
$E_{\rm max}^{{\rm AB}, a}$ & $[(\mu^2 L^2)(\mu\hbar c)]^{1/4}$ & $(\mu \hbar c)^{1/2}$  & $(\mu \hbar c)^{1/2}$\\
\hline
\end{tabular}
\caption{AB variable definitions}\label{table:AB_vars}
\end{table}

\subsection{Indirect Coupling}
Now, we present the spectrum of neutrinos indirectly emitted from cosmic string loops. More concretely, we first present the spectrum of real scalar particles directly emitted from string loops with quasi-cusps, quasi-kinks, and kink-kink collisions. Then, we write the spectrum of neutrinos emitted from a real scalar.

Once again, there are no obvious suppressions, and so the string feature spectra and emitted power are
\begin{align}
\frac{dN^\phi_a}{dE_\phi} &= \tilde{\Gamma}_a^\phi\left(\frac{E_\phi L}{\hbar c}\right)^{q_a^\phi}\frac{\mu \hbar c}{E_\phi^3},\\
P_a^\phi &= \frac{\Gamma_a^\phi \mu c}{(L/\ell_\phi)^{p_a^{\phi}}},
\end{align}
 with $E_\phi$ the lab frame energy of the $\phi$ particle (different from the neutrino energy $E$) and $\ell_\phi = \ell_{\rm Yu} = \ell_{\rm casc} = \lambda_\phi$. All other variable definitions are placed in Table~\ref{table:phi_vars}. After the real scalar is emitted, we assume it emits neutrinos instantaneously. 
\begin{table}[h!]
\begin{tabular}{|c|c|c|c|}
\hline
$a$ & $qc$ & $qk$ & $kk$\\
\hline
$q_a^{\phi}$ & $2/3$ & $1/3$ & $0$\\
$p_a^{\phi}$ & $1/2$ & $1$  & $1$\\
$\tilde{\Gamma}^\phi_a$ & $\alpha^2 \mathcal{S}_{qc}^\phi\Theta^2/[2(2\pi)^2]$ & $\alpha^2 \mathcal{S}_{qk}^\phi \Theta/[2(2\pi)^2]$ & $\alpha^2 \mathcal{S}_{kk}^\phi/(2\pi)^2$\\
$\Gamma^{\phi}_a$ & $6\psi^{-1/3}\tilde{\Gamma}^{\phi}_{qc}$ & $6\psi^{-2/3}\tilde{\Gamma}_{qk}^{\phi}$  & $2\tilde{\Gamma}_{kk}^{\phi}$\\
$E_{\rm min}^{\phi, a}$ & $\psi m_\phi c^2\sqrt{m_\phi cL/\hbar}$ & $\psi m_\phi c^2\sqrt{m_\phi cL/\hbar}$  & $m_\phi c^2$\\
$E_{\rm max}^{\phi, a}$ & $[(\mu^2 L^2)(\mu\hbar c)]^{1/4}$ & $(\mu \hbar c)^{1/2}$  & $(\mu \hbar c)^{1/2}$\\
\hline
\end{tabular}
\caption{Variable definitions for the real scalar $\phi$}\label{table:phi_vars}
\end{table}

\subsubsection{Yukawa}
Through a Yukawa coupling, two neutrinos are emitted from the heavy real scalar $\phi$  with an isotropic (i.e flat energy) spectrum
\begin{align}
\frac{dN^{\rm Yu}}{dE} &= \frac{1}{E_\phi},
\end{align}
with $m_\nu \ll E  \leq E_\phi$.

Therefore, the total number of neutrinos emitted from a cosmic string loop is also independent of the neutrino energy,
\begin{align}
\frac{dN_a^{\rm Yu}}{dE} &= \tilde{\Gamma}^{\rm Yu}_a\left(\frac{\lambda_\phi}{L}\right)^{q_a^{\rm Yu}}\frac{\mu \lambda_\phi^2}{\hbar c}\frac{1}{m_\phi c^2},
\end{align}
with all variable definitions in Table~\ref{table:Yu_vars}.
\begin{table}[h!]
\begin{tabular}{|c|c|c|c|}
\hline
$a$ & $qc$ & $qk$ & $kk$\\
\hline
$q_a^{\rm Yu}$ & $1/2$ & $0$ & $0$\\
$\tilde{\Gamma}^{\rm Yu}_a$ & $(3/7)\psi^{-7/3}\tilde{\Gamma}^{\phi}_{qc}$ & $(3/8)\psi^{-8/3}\tilde{\Gamma}^{\phi}_{qk}$  & $(1/3)\tilde{\Gamma}_{kk}^{\phi}$\\
$E_{\rm min}^{\rm Yu, a}$ & $m_\nu c^2$ & $m_\nu c^2$  & $m_\nu c^2$\\
$E_{\rm max}^{\rm Yu, a}$ & $E_{\rm max}^{\phi, qc}$ & $E_{\rm max}^{\phi, qk}$  & $E_{\rm max}^{\phi, kk}$\\
\hline
\end{tabular}
\caption{Yukawa variable definitions}\label{table:Yu_vars}
\end{table} 
\subsubsection{Cascade}
After the heavy scalar decays, a cascade of particles decays ensues, of which neutrinos are one of the end products. In according with previous studies~\cite{hep-ph/0009053, hep-ph/0108098, hep-ph/0211406, hep-ph/0307279}, we assume that the decay spectra follows a power law with index $\sim -2$ and that approximately all of the energy is transferred to pions, which then decay to give half of their energy to neutrinos. After imposing conservation of energy in the decay between neutrinos and the heavy real scalar we obtain  
\begin{align}
\frac{dN^{\rm casc}}{dE} &= \frac{b_*}{2}\frac{E_\phi}{E^2},
\end{align}
 with $b_* = \log\left(E_{\rm max}^{\rm casc}/E_{\rm min}^{\rm casc}\right)^{-1}$. As a result, the total number of neutrinos emitted from a cosmic string loop is
\begin{align}
\frac{dN^{\rm casc}_a}{dE} &= \tilde{\Gamma}^{\rm casc}_a b_*\left(\frac{\lambda_\phi}{L}\right)^{q^{\rm casc}_a}\frac{\mu \lambda_\phi}{E^2},
\end{align}
with all variable definitions in Table~\ref{table:casc_vars}.
\begin{table}[h!]
\begin{tabular}{|c|c|c|c|}
\hline
$a$ & $qc$ & $qk$ & $kk$\\
\hline
$q_a^{\rm casc}$ & $-1/2$ & $0$ & $0$\\
$\Gamma^{\rm casc}_a$ & $(1/4)\Gamma^{\phi}_{qc}$ & $(1/4)\Gamma^{\phi}_{qk}$  & $(1/4)\Gamma_{kk}^{\phi}$\\
$E_{\rm min}^{\rm casc, a}$ & $(1/2)\sqrt{m_\phi c^2 Q_h}$ & $(1/2)\sqrt{m_\phi c^2 Q_h}$  & $(1/2)\sqrt{m_\phi c^2 Q_h}$\\
$E_{\rm max}^{\rm casc, a}$ & $0.1 E_{\rm max}^{\phi, qc}$ & $0.1 E_{\rm max}^{\phi, qk}$  & $0.1 E_{\rm max}^{\phi, kk}$\\
\hline
\end{tabular}
\caption{Cascade variable definitions, with $Q_{h} = 1\ {\rm GeV}$ the hadronization energy scale.}\label{table:casc_vars}
\end{table}

\section{String Loop Population}\label{sec:string_pop}
A loop of initial length $L_i$ at time $t_i$ will contract as it radiates energy from various string features. For the string interaction models presented here, this energy may either be in the form of gravitational waves, neutrinos, or real scalar fields. However, we do not consider emission via all these channels at once. Instead, in order to determine the evolution of the loop distribution function, we consider emission in a pair of channels: first, from gravitational waves and second, from a single specified particle model. This choice is done because cosmic string loops are always expected to radiate gravitationally and our models are an addition beyond the standard framework. As a result, the center of mass energy $\mu L$ of a loop decrease over time according to
\begin{align}\label{eq:loop_radiation}
\mu\frac{dL}{dt} &= -\Gamma_g G\mu^2 c^{-3} - P_a^e,
\end{align} 
with $\Gamma_g \in [50, 100]$. The first term encodes loop emission of gravitational waves, while the second term specifies the emission $e$ from string feature $a$. Moreover, loops with length $L > L_a^e = \ell_e \left[(\Gamma_a^e/\Gamma_g)/(G\mu c^{-4})\right]^{1/p_a^e}$ emit more energy in the form of gravitational waves than from emission $e$ from string feature $a$. 

In general, Eq.\eqref{eq:loop_radiation} does not have an analytic solution for arbitrary initial loop length. However, loops with $L_i < L_a^e$ will always emit more particles than gravitational waves, and those with $L_i \gg L_a^e$ more gravitational waves than particles. Therefore we solve for the evolution of loop length with these two conditions. Moreover, in practice, the conditon $L_i \gg L_a^e$ is relaxed to $L_i > L_a^e$, so that there are only two regimes:
\begin{align}
L(t_i, t, L_i) &= \left[L_i^{1 + p_a^e} - \left(L_{\rm min}^{e, a}\right)^{1  + p_a^e}\right]^{\frac{1}{1 + p_a^e}}\Theta\left(L_a^e - L_i\right)\\
\nonumber &+ \left[L_i - \Gamma_g G \mu c^{-3}(t - t_i)\right]\Theta\left(L_i - L_a^e\right),
\end{align}
which can be piecewise-inverted to solve for $L_i$ as a function of $L$. Here, $L_{\rm min}^{e, a} = \left[(1 + p_a^e)\Gamma_a^e c(t - t_i)\ell_e^{p_a^e}\right]^{1/(1 + p_a^e)}$.

\begin{figure}
\includegraphics[width = 0.55\textwidth]{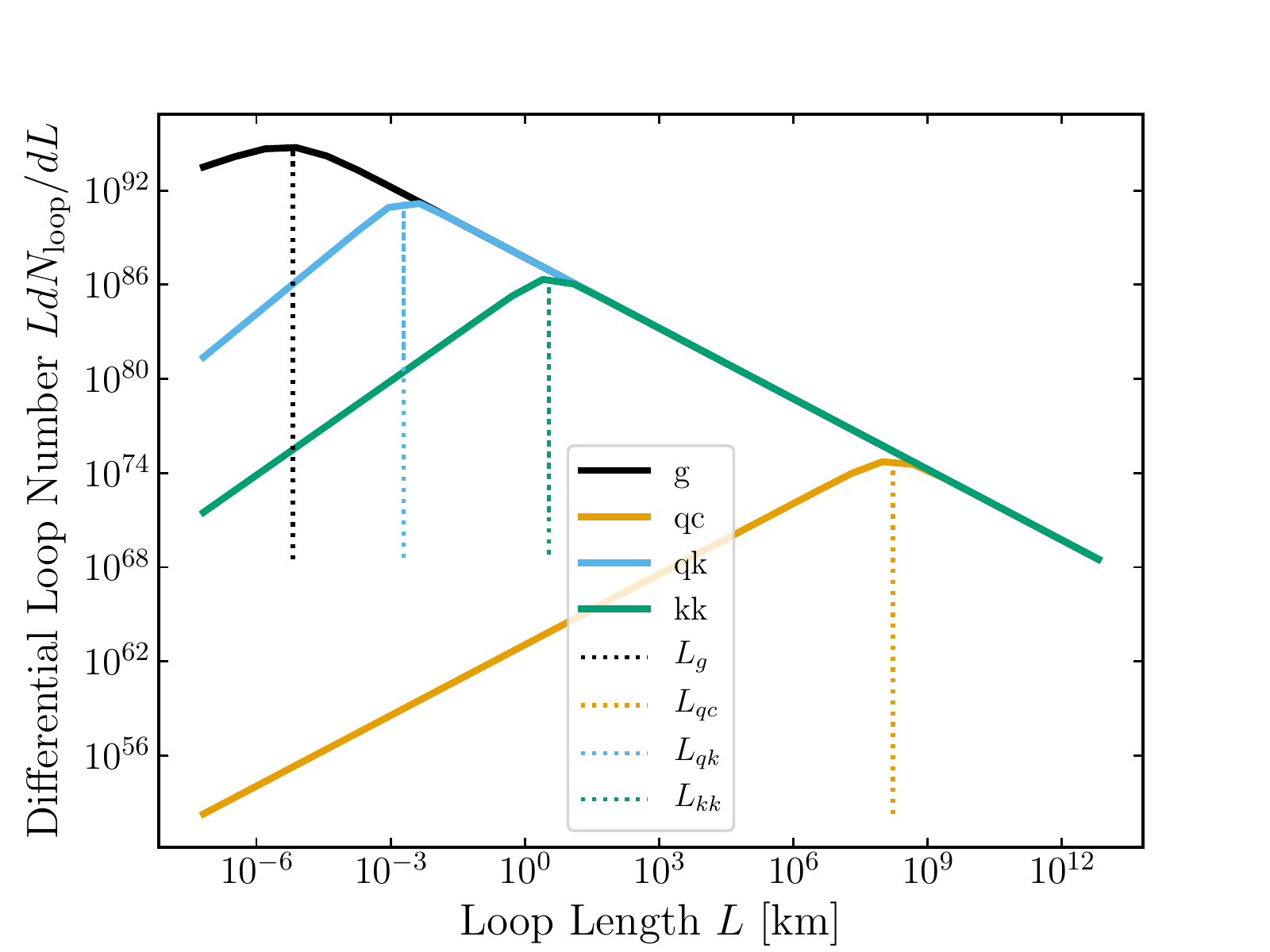}
\caption{The differential loop number $dN_{\rm loop}/dL = \chi^3 dn_{\rm loop}/dL$, with $\chi$ the size of the comoving horizon, evaluated at $z = 0$. The solid black line is the number assuming only gravitational emission, while the solid orange (blue) [green] line is due to both gravitational emission and AB emission from quasi-cusps (quasi-kinks) [kink-kink collisions]. The vertical dotted lines indicate the length $L_{\rm min}^{e, a}$.}\label{fig:dndL}
\end{figure}

\begin{figure}
\includegraphics[width = 0.55\textwidth]{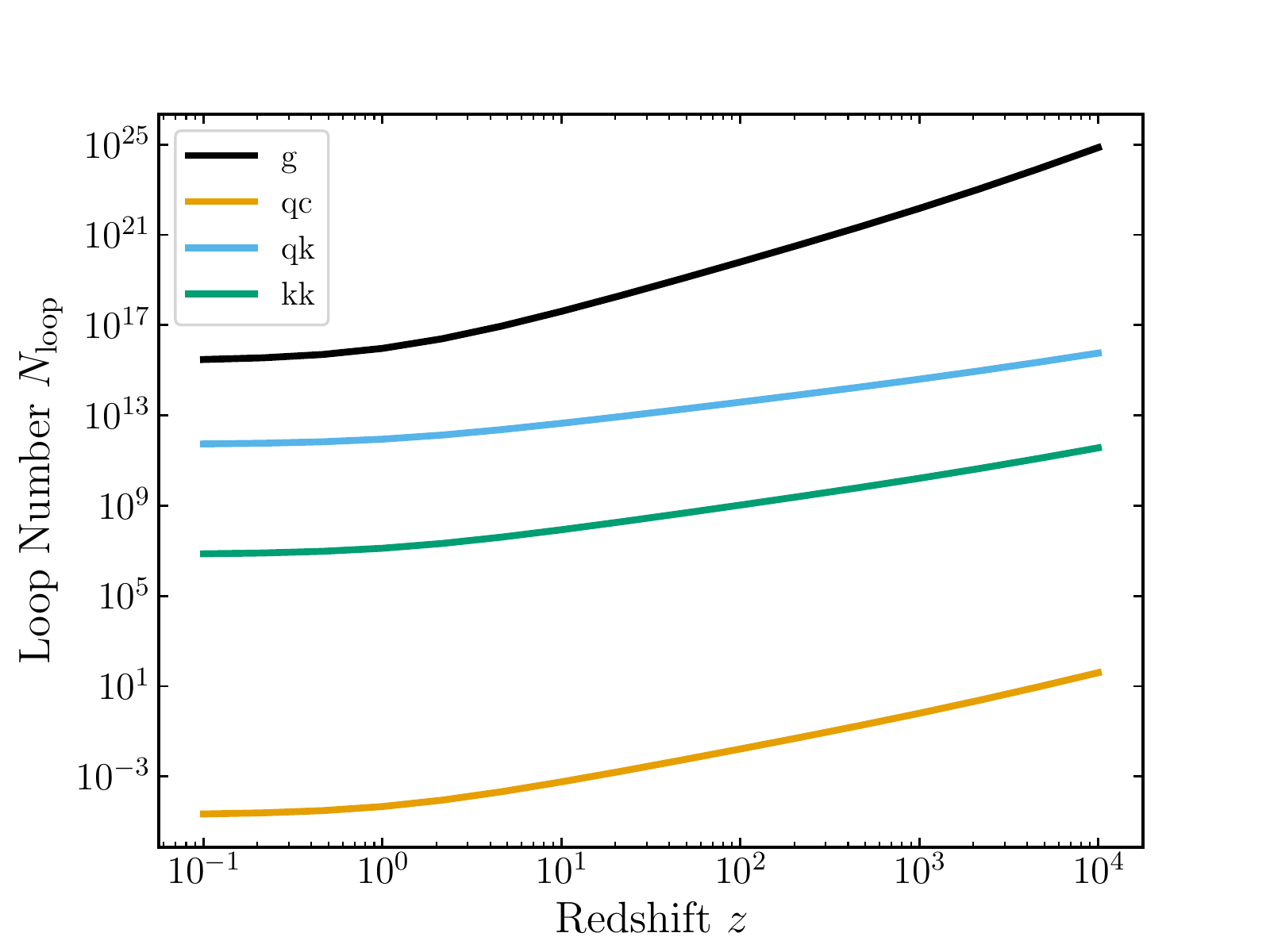}
\caption{The number of loops $N_{\rm loop} = \chi^3 n_{\rm loop}$, with $\chi$ the size of the comoving horizon, as a function of redshift. The label and color scheme follows that of Fig.~\ref{fig:dndL}. Hence, loop distributions with smaller $L_{\rm min}$ have higher numbers.}\label{fig:n}
\end{figure}

While some cosmic string loops are present at the initial $U(1)$ phase transition, most are formed after string segments intersect and commute, breaking off into smaller loops. Here, we assume this string loop population has relaxed to a steady-state self-similar solution. As a result, we neglect terms that involve string collision and string self-interactions. While these loops are produced both during periods of radiation and matter domination, those produced during matter domination are less abundant~\cite{hep-ph/9803414}. Therefore, we write the loop distribution as $dn^{\rm loop}/dL = dn_r^{\rm loop}/dL$, with 
\begin{align}
\frac{dn_r^{\rm loop}(t, L)}{dL}  &=
\frac{\zeta_r}{2}\frac{a_{\rm eq}^3}{[a(t_{\rm eq})\chi(t_{\rm eq})]^{3/2}L_0^{5/2}}\left(\frac{L}{L_0}\right)^p\\
\nonumber&\times \begin{cases}\Theta\left(\beta_r - \frac{L}{2ct} \right)& t\leq t_{\rm eq}\\
\Theta\left(\beta_r - \frac{ L_{eq}}{2ct_{\rm eq}}\right)\ & t > t_{\rm eq}
\end{cases}
\end{align} 
the distribution of loops created during radiation-domination at a time $t$. Moreover, $t_{\rm eq}$ is the time of matter-radiation equality, $\chi$ the comoving horizon distance, $\zeta_r = 1.04$ a normalization factor, $\beta_r = 0.05$ the typical scale of loops produced radiation domination relative to the size of the horizon. Finally, $L_0 = L_i(0, t, L)$ and $L_{\rm eq} = L_i(t_{\rm eq}, t, L)$ are the lengths of a loop at $t = 0$ and $t_{\rm eq}$. We show some example distributions for cosmic string loops in Fig.~\ref{fig:dndL} and Fig.~\ref{fig:n}.

\section{Results}\label{sec:results}
Given the emission spectra of neutrinos from a single cosmic string loop, as well as the distribution of cosmic string loops, we now present both the source function and flux for each phenomenological case. We then use the dominant flux to place an upper bound on the fraction $f_a^e$ of HEAN sourced by cosmic string loops. The bounds are obtained in the following manner.

First, evaluating Eq.~\eqref{eq:gen_source}, we obtain
\begin{align}\label{eq:spec_source}
S_{a}^e(t, E) &= c A_a^e \frac{dN^e_a[E, L_{\rm min}^{e, a}(t)]}{dE}\frac{dn_{\rm loop}[0, L_{\rm min}^{e, a}(t)]}{dL},\\
A_a^e &\equiv \frac{4}{1 + p_a^e}\frac{\Gamma\left(\frac{5 + 2q_a^e}{2 + 2p_a^e}\right)}{\Gamma\left(\frac{7 + 2p_a^e + 2q_a^e}{2 + 2p_a^e}\right)} \\
&\nonumber\times_2F_1\left[1 + \frac{3}{2 + 2p_a^e}, \frac{5 + 2q_a^e}{2 + 2p_a^e}, \frac{7 + 2p_a^e + 2q_a^e}{2 + 2p_a^e}, -1\right]
\end{align} 
with $\Gamma(n)$ the Gamma function, and $_2 F_1(a, b, c, d)$ a hypergeometric function. In this expression, we remind that $a \in \{qc, qk, kk\}$ and $e \in \{(2), {\rm AB}, {\rm Yu}, {\rm casc}\}$.

We define the index of the local energy spectrum through the expression $dN_a^e/dE \propto E^{-\gamma_a^e}$. Using Eq.~\eqref{eq:Dnu} and Eq.~\eqref{eq:spec_source}, we evaluate Eq.~\eqref{eq:boltzmann}, after changing variables from time to redshift via $dt/dz = -1/[H(z)(1 + z)]$, to obtain
\begin{align}\label{eq:final_spec}
\Phi_a^e(t, E) &= I_a^e(t, E)\frac{c^2}{H_0}\frac{dN_a^e[E, L_{\rm min}^{e, a}(t)]}{dE}\frac{dn_{\rm loop}[0, L_{\rm min}^{e, a}(t)]}{dL}\\
I_a^e(t, E) &\equiv A_a^e \int_{z(t)}^{z_\nu(t, E)} \frac{dz}{E(z)}(1 + z)^{-\gamma_a^e}f(z)^{-(q_a^e + \frac{5}{2})/(p_a^e + 1)},
\end{align} 
with $H(z) = H_0 E(z)$ the Hubble parameter, $H_0$ Hubble's constant, $E(z) = \left[\Omega_m(1 + z)^3 + (1 - \Omega_m) + \Omega_r(1 + z)^4\right]^{1/2}$ for $\Lambda{\rm CDM}$, $\Omega_m$ the matter-density parameter, $\Omega_r$ the radiation-density parameter , and $f(z) = t(z)/t$. We present the values for $I_a^e(t_0, E_{\rm min})$ in Table~\ref{table:I} using Planck 2018 parameters~\cite{1807.06209}. 

\begin{table}[h!]
\begin{tabular}{|c|c|c|c|}
\hline
$I_a^e$ & $qc$ & $qk$ & $kk$\\
\hline
$(2)$ & N/A & N/A & $4.35 \times 10^{10}$\\ 
AB & $94900$ & $15.2$ & $241$\\ 
Yu & $3.25\times 10^{11}$ & $2.40\times 10^6$ & $2.40\times 10^6$\\ 
casc & $2.56$ & $2.04$ & $2.04$\\ 
\hline
\end{tabular}
\caption{Tabulated values for $I_a^e(t_0, E_{\rm min})$, $E_{\rm min} = 16\ {\rm TeV}$, with $a$ specified by the column and $e$ by the row. For emission of type $(2)$, quasi-cusps and -kinks are helicity suppressed and so we do not consider them here.}\label{table:I}

\end{table}

\begin{figure}
\includegraphics[width = 0.55\textwidth]{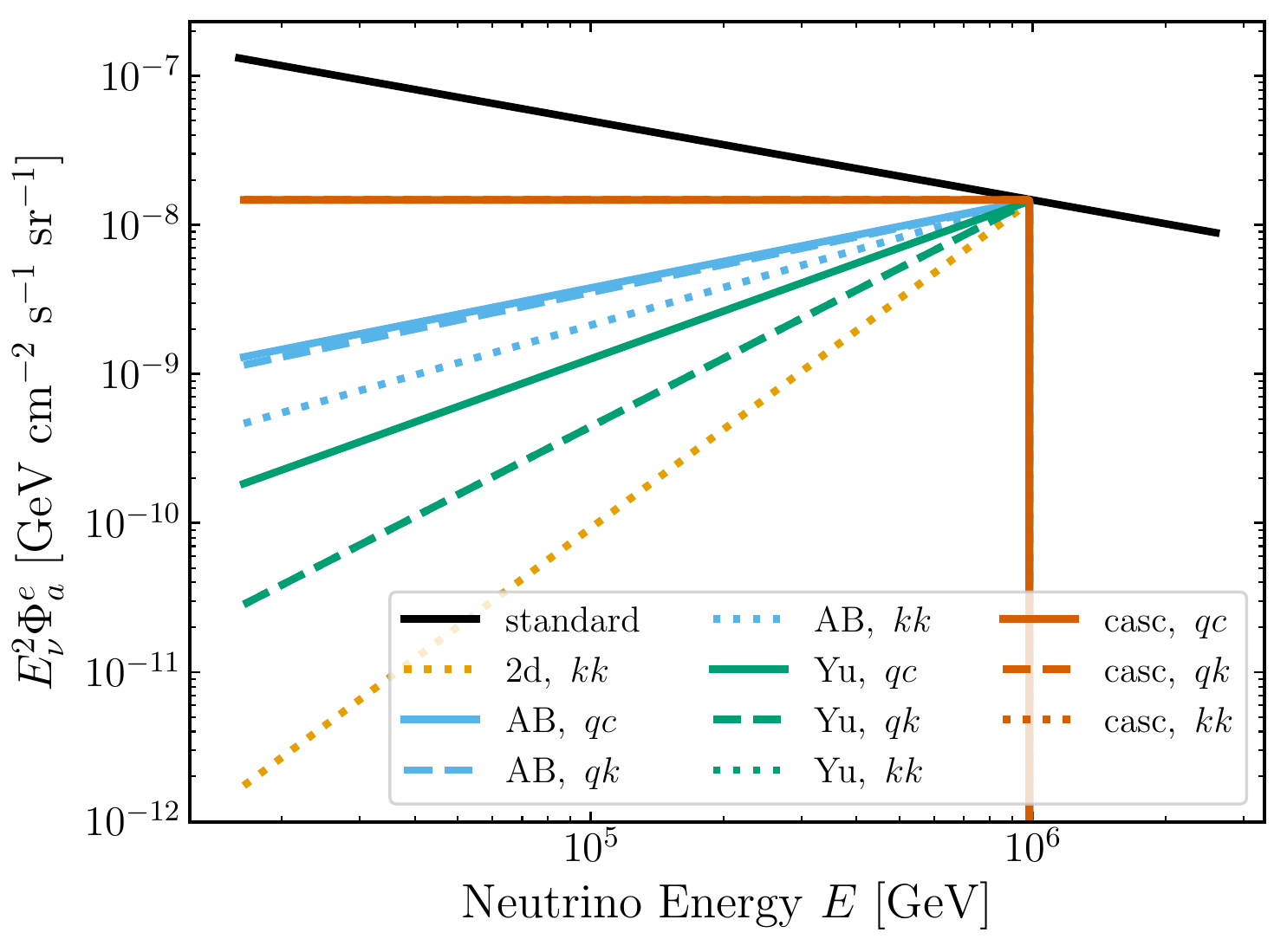}
\caption{Spectra of HEAN emitted from cosmic strings using Eq.~\eqref{eq:maineq} compared to the observed HEAN spectrum (in solid black) using Eq.~\eqref{eq:obs_spec}. The orange (blue) [green] \{red\} line indicates HEAN emission via the $(2)$ (AB) [Yu] \{casc\} model. Moreover, solid (dashed) [dotted] lines indicate that the string population contains quasi-cusps (quasi-kinks) [kink-kink collisions].
We choose $E_{\rm max}^{e, a} = 10^{6}\ {\rm GeV}$. For large enough amplitude values, the spectrum may appear as a bump before the sharp cutoff.
}\label{fig:nu_spec}
\end{figure}

To make easy connection with observation, we reparametrize the neutrino spectrum $\Phi_a^e(t_0, E)$ today as a power law with a sharp cutoff,
\begin{align}\label{eq:maineq}
\Phi_a^e(t_0, E) \simeq C_0 B_a^e (E/E_0)^{-\beta_a^e}\Theta\left(E - E_{\rm max}^{e,a}\right)\Theta\left(E- E_{\rm min}^{e, a}\right),
\end{align}
with $C_0 = 2\times 10^{-18}\ {\rm GeV}^{-1}{\rm cm}^{-2}{\rm s}^{-1}{\rm sr}^{-1}$ and $E_0 = 100\ {\rm TeV}$. Note that, for most cosmic string parameter values, $E_{\rm min}^{a, e}$ is much smaller than observed HEAN energies and so the low-energy cutoff can be ignored. We write this equation as an approximate equality as the spectral index $\beta_a^e$ has a nonzero running with energy, $d\beta_a^e/dE \neq 0$. However, this running is small and so we average its value over the observed energy range. Moreover, note that $\beta_a^e\neq \gamma_a^e$ as the energy dependence of the neutrino horizon shifts the spectral index, which we show in Table~\ref{table:beta}. We show the dependence of the amplitude $B_a^e$ on our model parameters in Table~\ref{table:B}. In order to save space in the table, we include a scaling of the $b_*$ parameter in Eq.~\eqref{eq:bstar}. Using the new parametrization of Eq.~\eqref{eq:maineq}, we plot some example spectra in Fig.~\ref{fig:nu_spec} 
\begin{table}[h!]
\begin{tabular}{|c|c|c|c|}
\hline
$\beta_a^e$ & $qc$ & $qk$ & $kk$\\
\hline
$(2)$ & N/A & N/A & $-0.27$\\ 
AB & $1.37$ & $1.38$ & $1.14$\\ 
Yu & $0.855$ & $0.439$ & $0.439$\\ 
casc & $2$ & $2$ & $2$\\ 
\hline
\end{tabular}
\caption{Tabulated values for $\beta_a^e$, with $a$ specified by the column and $e$ by the row.}\label{table:beta}
\end{table}

\begin{table*}[ht!]
\scalebox{0.82}{
\begin{tabular}{|c|c|c|c|}
\hline
$B_a^e$ & $qc$ & $qk$ & $kk$\\
\hline
$(2)$ & N/A & N/A & $1.09\times 10^{-4}\left(\frac{\tilde{\Gamma}_{kk}^{(2)}}{10^{36}}\right)^{-1/4}\left(\frac{G \mu c^4}{4.5 \times 10^{-26}}\right)^{-3/8}$\\ 
AB & $0.383\left(\frac{\tilde{\Gamma}^{\rm AB}_{qc}}{10^{-25}}\right)^{-2/3}\left(\frac{G\mu c^{-4}}{4.5\times 10^{-26}}\right)^{5/12}$ & $2.71\left(\frac{\tilde{\Gamma}^{\rm AB}_{qk}}{10^6}\right)^{-3/14}\left(\frac{G\mu c^{-4}}{4.5\times 10^{-26}}\right)^{17/21}$ & $0.0212\left(\frac{\tilde{\Gamma}_{kk}^{\rm AB}}{10^{10}}\right)^{-1/4}\left(\frac{G\mu c^{-4}}{4.5\times 10^{-26}}\right)^{5/8}$\\ 
Yu & $0.569\left(\frac{\tilde{\Gamma}^{\rm Yu}_{qc}}{10^{-28}}\right)^{-1}\left(\frac{G\mu c^{-4}}{4.5\times 10^{26}}\right)\left(\frac{m_\phi c^2}{10^5\ {\rm GeV}}\right)^{-5/2}$ & $9.28\left(\frac{\tilde{\Gamma}^{\rm Yu}_{qk}}{10^{10}}\right)^{-1/4}\left(\frac{G\mu c^{-4}}{4.5\times 10^{26}}\right)\left(\frac{m_\phi c^2}{10^5\ {\rm GeV}}\right)^{-7/4}$ & $62.8\left(\frac{\tilde{\Gamma}^{\rm Yu}_{kk}}{10^{20}}\right)^{-1/4}\left(\frac{G\mu c^{-4}}{4.5\times 10^{-26}}\right)\left(\frac{m_\phi c^2}{10^{5}\ {\rm GeV}}\right)^{-7/4}$\\ 
casc & $\frac{3.42}{1000}\left(\frac{b_*}{7.4}\right)\left(\tilde{\Gamma}^{\rm casc}_{qc}\right)^{-1/3}\left(\frac{G\mu c^{-4}}{4.5\times 10^{-24}}\right)\left(\frac{m_\phi c^2}{10^7\ {\rm GeV}}\right)^{1/6}$ & $28.2\left(\frac{b_*}{7.4}\right)\left(\tilde{\Gamma}^{\rm casc}_{qk}\right)^{-1/4}\left(\frac{G\mu c^{-4}}{4.6\times 10^{-24}}\right)\left(\frac{m_\phi c^2}{10^7\ {\rm GeV}}\right)^{1/4}$ & $307 \left(\frac{b_*}{7.4}\right)\left(\tilde{\Gamma}_{kk}^{\rm casc}\right)^{-1/4}\left(\frac{G \mu c^{-4}}{4.5 \times 10^{-24}}\right)\left(\frac{m_\phi c^2}{10^7\ {\rm GeV}}\right)^{1/4}$\\ 
\hline
\end{tabular}}
\caption{Tabulated values for $B_a^e$, with $a$ specified by the column and $e$ by the row. The scaling of $b_*$ is shown in Eq.~\eqref{eq:bstar}. Fiducial values are chosen so that they are not ruled out by HEAN spectra observations.}\label{table:B}
\end{table*}
 
\begin{align}\label{eq:bstar}
\exp\left(b_*\right) &= 1640\left(\frac{G\mu c^{-4}}{4.5\times 10^{-24}}\right)^{1/2}\left(\frac{m_\phi c^2}{10^7\ {\rm GeV}}\right)^{1/2}
\end{align}

\begin{figure}
\includegraphics[width = 0.55\textwidth]{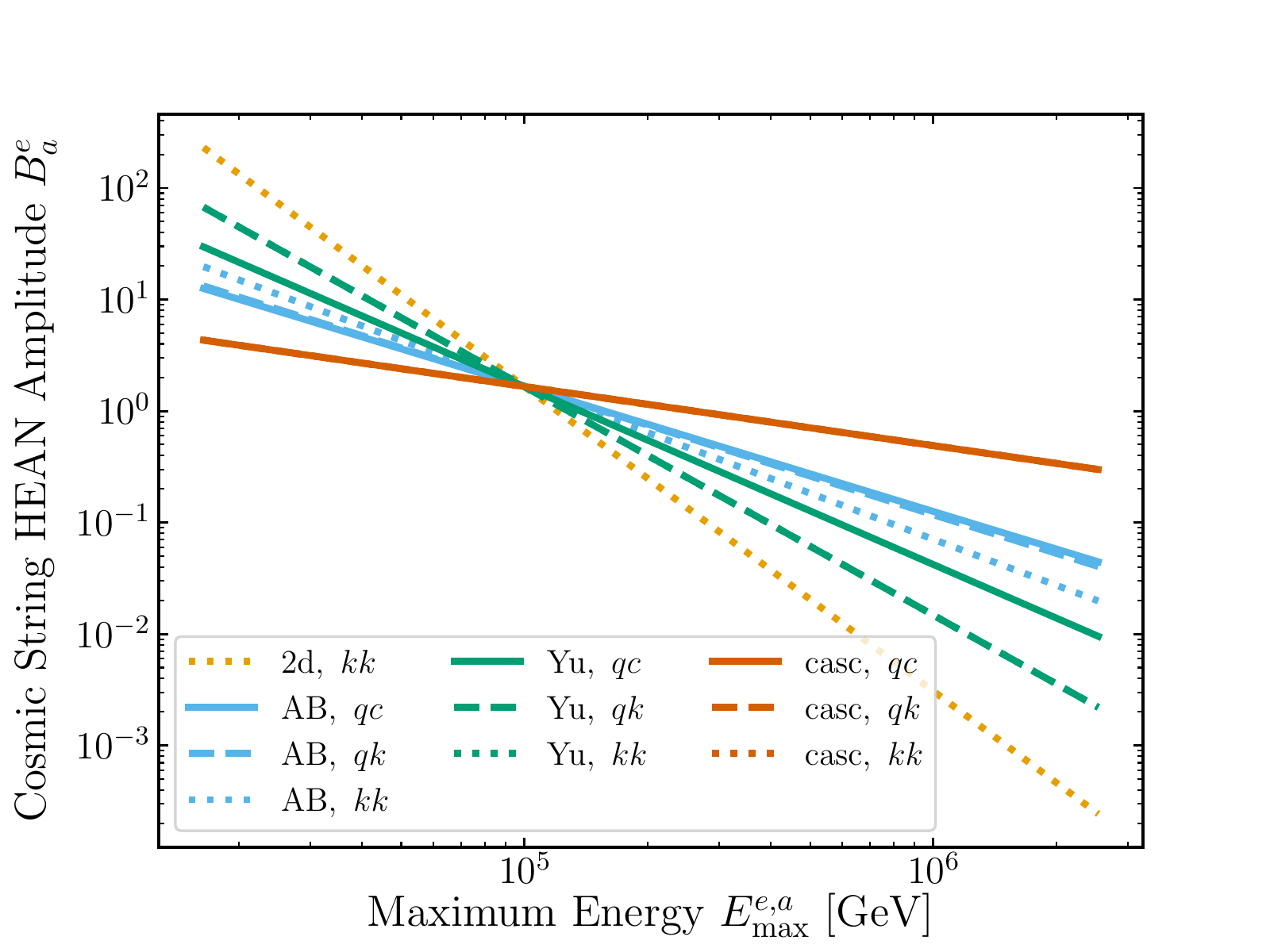}
\caption{The maximum amplitude $B_a^e$ of HEANs that come from a population of cosmic string loops using Eq.~\eqref{eq:BCS}. The orange (blue) [green] \{red\} line indicates HEAN emission via the $(2)$ (AB) [Yu] \{casc\} model. Moreover, solid (dashed) [dotted] lines indicate that the string population contains quasi-cusps (quasi-kinks) [kink-kink collisions]. All lines intersect at $E_{\rm max}^{e, a} = E_0$ by construction of our parameterization. Values of $B_a^e$ above a given line are ruled out. Table~\ref{table:B} translates these amplitudes into cosmic string parameters. 
}\label{fig:BCS}
\end{figure}

We now identify the viable parameter space of cosmic string HEAN emission subject to the constraint that it not greater than the observed HEAN spectrum, $\Phi_a^e(E) \leq \Phi_{\rm HEAN}(E)$, for all energies. We model the observed HEAN spectrum as a power law with spectral index $\gamma = 2.53$~\cite{1907.11266},
\begin{align}\label{eq:obs_spec}
\Phi_{\rm HEAN}(E) &= C_0 \Phi_0 \left(E/E_0\right)^{-\gamma},
\end{align} 
with $\Phi_0 = 1.66$. We take the observed HEAN energy range to be $E_{\rm min} = 16\ {\rm TeV} \leq E \leq E_{\rm max} = 2.6\ {\rm PeV}$. As a result, the three equations
\begin{align}\label{eq:BCS}
B_a^e \leq \Phi_0 \left(E_{\rm max}^{e,a}/E_0\right)^{\beta_a^e - \gamma},\\
E_{\rm min} \leq E_{\rm max}^{e, a} \leq E_{\rm max},\\
E_{\rm min}^{e, a} \leq E_{\rm max}^{e, a}
\end{align} 
define a region in the cosmic string parameter space that is viable to contribute to the HEAN flux and whose upper bound we show in Fig.~\ref{fig:BCS}. Parameters that are above this upper bound are ruled out, as they would lead to a HEAN spectrum larger than what we observe. In order to relate these equations to the original parameters, one can use the formulas listed in Tables~\ref{table:beta} and~\ref{table:B}, along with the definition of $E_{\rm max}^{e, a}$ listed in the Tables in Sec~\ref{sec:emission}. 

\begin{figure}
\includegraphics[width = 0.55\textwidth]{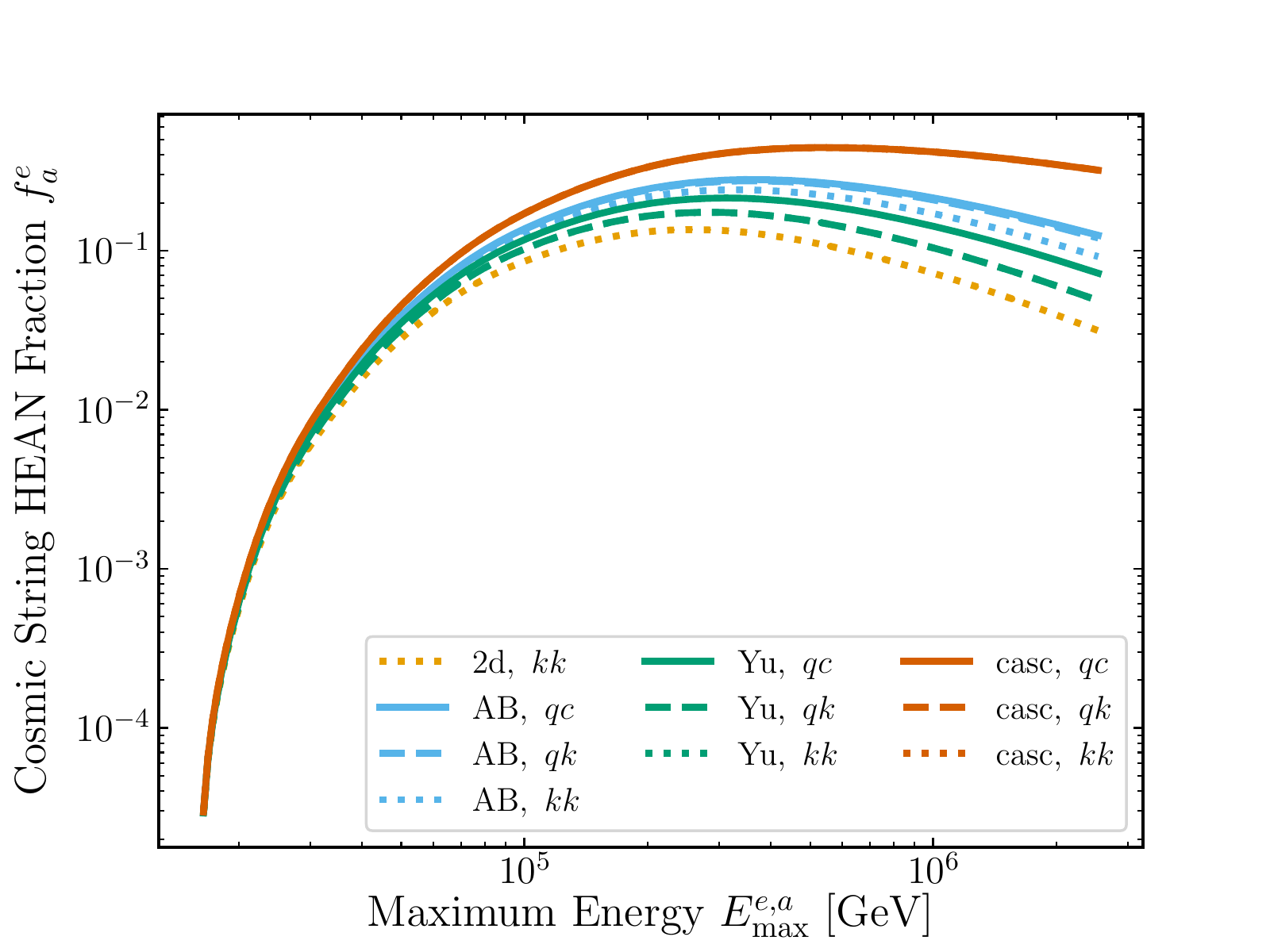}
\caption{The maximum fraction of HEANs that come from a population of cosmic string loops using Eq.~\eqref{eq:fCS}. The orange (blue) [green] \{red\} line indicates HEAN emission via the $(2)$ (AB) [Yu] \{casc\} model. Moreover, solid (dashed) [dotted] lines indicate that the string population contains quasi-cusps (quasi-kinks) [kink-kink collisions].   
}\label{fig:fCS}
\end{figure}

The fraction $f_a^e$ of observed neutrinos associated with a cosmic string spectrum given by emission model $e$ and string feature $a$ is then
\begin{align}\label{eq:fCS}
f_a^e= \frac{\int_{E_{\rm min}}^{E_{\rm max}}dE A_{\rm eff}(E) \Phi_a^e(t_0, E)}{\int_{E_{\rm min}}^{E_{\rm max}} dE A_{\rm eff}(E) \Phi_{\rm HEAN}(E)},
\end{align} 
with $A_{\rm eff}(E)$ the effective area of IceCube for muon neutrinos, which we take from Ref.~\cite{1311.5238}. We plot the maximum contribution of cosmic string loops [i.e. when $\Phi_a^e(t_0, E^{e, a}_{\rm max}) = \Phi_{\rm HEAN}(E_{\rm max}^{e, a})$]in Fig.~\ref{fig:fCS}. 
\section{Discussion}\label{sec:disc}
We clarify five assumptions and present six comments. First, in each model of neutrino emission we assume a single neutrino. If there are multiple neutrino species coupled to the string, then energy extracted from the loop will be a sum over all neutrino emission channels. Moreover, since  cosmic strings are distant objects, then the observed spectrum of neutrinos will be a sum of the spectrum of each neutrino channel weighed by the corresponding probability of oscillating into that neutrino. Note that even if there is only one neutrino coupled, then flavor oscillations will decrease the spectral amplitude in that flavor. Regardless, the effects of oscillations can always be absorbed into a redefintion of $\tilde{\Gamma}$ and so our results can be scaled appropriately to include them. 

Second, if the neutrino is a Majorana fermion instead of a Dirac fermion, then $\bar{\nu} = \nu^{\rm T}\mathcal{C}$ with $\mathcal{C}$ the charge conjugation matrix. This replacement will not change the spectral index, and should not change the amplitude of emission by more than an $\mathcal{O}(1)$ coefficient. 

Third, for the indirect emission models, we assume the heavy real scalar instantaneously decays into neutrinos.

Fourth, we did not consider cosmic string loop populations with multiple features (e.g. loops that have both quasi-kinks and -cusps). Since cusps extract more energy from the string than kinks and quasi-kinks, but typically have smaller amplitudes, we expect that the presence of cusps would decrease the expected amplitude in the HEAN energy range (or alternatively, the presence of kinks to increase the amplitude). Thus, our results safely represent an upper limit on the possible contribution of cosmic strings to the HEAN spectrum.  

Fifth, we assume that the population of cosmic string is characterized by a single string tension value. Instead, it is possible that there exists multiple varieties of cosmic strings in the Universe, with each cosmic string characterized by a different string tension, and thus the resulting spectrum would be the sum of these two types of strings. In addition, the string tension may have some time dependence~\cite{hep-ph/0503227}, leading to a HEAN spectrum that would be average over the distribution of tension values. Both of these cases are beyond the scope of this work. 

While we do consider a wide variety of emission models here, the list is not exhaustive. For example, we did not consider 2-body emission of real scalars from cosmic strings than then decay in HEAN. In the case of 2-body emission of real scalars, this model would not change the spectrum index of emission relative to its 1-body counterpart. This similarily is because the index is controlled by the Yukawa and cascade decays. Therefore, while the precise values for the amplitude $\tilde{\Gamma}$ may change, the maximum contribution to the HEAN spectrum will not. In other cases, unless the spectral index of emission just so happens to match the one of the observed spectrum, we expect our limit, of no more than $45\%$ of HEANs to come from cosmic strings, to hold.

In addition to emission models, it is also possible that cosmic strings collide and annihilate with one another into neutrinos. However, cosmic strings are very thin and so their annihilation cross-section is very small. Thus, we do not expect such a process to contribute greatly. 

Even though we find that the models presented are a subdominant portion of the total spectrum, the presence of a sharp cutoff implies that HEANs from cosmic strings may present as a distinct bump in the observed HEAN spectrum, opening up the possibility for their detection. Moreover, if cosmic strings exist, their gravitational wells would alter energies of traversing photons. Hence, in principle, cross correlations of HEAN maps with the cosmic microwave background would be able to distinguish cosmic strings from other subdominant contributions, although we expect such a signal to be very small. 

In each of our plots in Sec.~\ref{sec:results}, the region to the right of the orange dashed line requires either values of the coupling constant or string feature parameters that are greater than $\mathcal{O}(1)$. It is both difficult to create such a theory and is at odds with the perturbative approach we took to calculating the spectra. Despite this, we leave this region in our plots as it may be the case other models with similar effective parameters and spectral indices are viable. 

Moreover, in these plots, we only consider the constraints on the effective parameters describing HEAN emission from cosmic strings. At higher neutrino energies, where current and future experiments like ANITA~\cite{2012.07945} and POEMMA~\cite{1902.04005} can observe neutrinos, there will be additional constraints. The future upgrade of IceCube-Gen2~\cite{2008.04323} will also allow detections of HEANs at lower energies, thus extending the range of our plots. In addition, in the cascade case, there will be an emission of gamma rays that go along with the neutrinos. Treatment of both of these effects are a work in progress and beyond the scope of this work.
 
Finally, we note that since we took an effective field theory approach to our problem, the parameter spaces we have identified may be constrained by once linked to a corresponding UV completion. However, it is not inconceivable that these UV completions will still have unconstrained parameter spaces for HEAN emission. Regardless, such an investigation is beyond the scope of this work.

\section{Conclusion}\label{sec:conc}
In this paper we quantified the possible contribution of cosmic strings to the HEAN spectrum for a wide variety of models. First, we presented the general formula for calculating neutrino emission from distant sources and updated the calculation for the HEAN optical depth compared to previous works on cosmic string emission. In doing so, we both employed a more accurate numerical approach and included all seven channels of Standard Model neutrino self-interactions.

Then, in order to classify possible models, we took an effective field theory approach and deliniated two avenues of HEAN production: direct and indirect. In direct emission, the cosmic string emits HEANs through a direct coupling of neutrinos to the cosmic string, while in indirect emission the cosmic string emits a particle which then decays into HEANs. For both direct and indirect emission we consider two models each. That is, we considered direct emission of HEANs via a two-body emission and a Aharonov-Bohm coupling. For indirect emission, we considered the emission of a heavy real scalar which then decays into HEANs either from a Yukawa coupling or through a hadronic cascade. Aside from the cascade case, all other calculations have not been done before. 

In addition to the particular cosmic string phenomenology, the energy spectrum of HEANs is also determined by the geometry of the string. In particular, efficient cosmic-string particle emission must come either from quasi-cusps, quasi-kinks, or kink-kink collisions on the string. Previous work has not considered emission from kink-kink collisions. Therefore, for each emission model and string feature, we then calculated the local energy spectrum of HEANs emitted from the cosmic string.  

Next, we calculated the distribution of cosmic string loops that emit both gravitational waves and a given neutrino emission model that specifies a string feature. These loops are created during radiation domination and then shrink as they emit energy. We note again that the shrinking due to non-gravitational emission has not been considered in previous works. In doing this calculation, we then also identified the dominant forms of energy emission in cosmic string loops and deliniated their corresponding regimes. 

With the local energy spectrum and cosmic-string loop distribution specified, we then calculated the HEAN energy spectrum today using the Boltzmann equation for each emission model and string feature and obtained a simple power law in with a sharp cutoff in Eq.~\eqref{eq:maineq}. With these spectra, we then required each one must be less than the observed HEAN spectrum. This requirement led us to identify and constrain the corresponding parameter space of HEAN emission. As a result, we found that, with the models presented, cosmic strings can contribute no more than $\sim 45\%$ of HEANs.  
\subsection*{Acknowledgments}
C.C.S acknowledges the support of the Bill and Melinda Gates Foundation. This work was supported at Johns Hopkins by NSF Grant No.\  1818899 and the Simons Foundation. This work was finalized at the Aspen Center for Physics, which is supported by National Science Foundation grant PHY-1607611.


\begin{thebibliography} {9}
%\cite{1907.11266}
\bibitem{1907.11266}
A.~Schneider [IceCube],
``Characterization of the Astrophysical Diffuse Neutrino Flux with IceCube High-Energy Starting Events,''
PoS \textbf{ICRC2019}, 1004 (2020)
%doi:10.22323/1.358.1004
[arXiv:1907.11266 [astro-ph.HE]].
%49 citations counted in INSPIRE as of 26 Jun 2021

%\cite{astro-ph/9701231}
\bibitem{astro-ph/9701231}
E.~Waxman and J.~N.~Bahcall,
``High-energy neutrinos from cosmological gamma-ray burst fireballs,''
Phys. Rev. Lett. \textbf{78}, 2292-2295 (1997)
%doi:10.1103/PhysRevLett.78.2292
[arXiv:astro-ph/9701231 [astro-ph]].
%1081 citations counted in INSPIRE as of 29 Jun 2021

%\cite{0907.2227}
\bibitem{0907.2227}
R.~Abbasi \textit{et al.} [IceCube],
``Search for muon neutrinos from Gamma-Ray Bursts with the IceCube neutrino telescope,''
Astrophys. J. \textbf{710}, 346-359 (2010)
%doi:10.1088/0004-637X/710/1/346
[arXiv:0907.2227 [astro-ph.HE]].
%109 citations counted in INSPIRE as of 29 Jun 2021

%\cite{1101.1448}
\bibitem{1101.1448}
R.~Abbasi \textit{et al.} [IceCube],
``Limits on Neutrino Emission from Gamma-Ray Bursts with the 40 String IceCube Detector,''
Phys. Rev. Lett. \textbf{106}, 141101 (2011)
%doi:10.1103/PhysRevLett.106.141101
[arXiv:1101.1448 [astro-ph.HE]].
%130 citations counted in INSPIRE as of 01 Jul 2021

%\cite{1204.4219}
\bibitem{1204.4219}
R.~Abbasi \textit{et al.} [IceCube],
``An absence of neutrinos associated with cosmic-ray acceleration in $\gamma$-ray bursts,''
Nature \textbf{484}, 351-353 (2012)
%doi:10.1038/nature11068
[arXiv:1204.4219 [astro-ph.HE]].
%349 citations counted in INSPIRE as of 01 Jul 2021

%\cite{1412.6510}
\bibitem{1412.6510}
M.~G.~Aartsen \textit{et al.} [IceCube],
``Search for Prompt Neutrino Emission from Gamma-Ray Bursts with IceCube,''
Astrophys. J. Lett. \textbf{805}, no.1, L5 (2015)
%doi:10.1088/2041-8205/805/1/L5
[arXiv:1412.6510 [astro-ph.HE]].
%150 citations counted in INSPIRE as of 01 Jul 2021

%\cite{1601.06484}
\bibitem{1601.06484}
M.~G.~Aartsen \textit{et al.} [IceCube],
``An All-Sky Search for Three Flavors of Neutrinos from Gamma-Ray Bursts with the IceCube Neutrino Observatory,''
Astrophys. J. \textbf{824}, no.2, 115 (2016)
%doi:10.3847/0004-637X/824/2/115
[arXiv:1601.06484 [astro-ph.HE]].
%101 citations counted in INSPIRE as of 01 Jul 2021

%\cite{1702.06868}
\bibitem{1702.06868}
M.~G.~Aartsen \textit{et al.} [IceCube],
``Extending the search for muon neutrinos coincident with gamma-ray bursts in IceCube data,''
Astrophys. J. \textbf{843}, no.2, 112 (2017)
%doi:10.3847/1538-4357/aa7569
[arXiv:1702.06868 [astro-ph.HE]].
%96 citations counted in INSPIRE as of 26 Jun 2021

%\cite{1711.03757}
\bibitem{1711.03757}
F.~Tavecchio, C.~Righi, A.~Capetti, P.~Grandi and G.~Ghisellini,
``High-energy neutrinos from FR0 radio-galaxies?,''
Mon. Not. Roy. Astron. Soc. \textbf{475}, no.4, 5529-5534 (2018)
%doi:10.1093/mnras/sty251
[arXiv:1711.03757 [astro-ph.HE]].
%9 citations counted in INSPIRE as of 26 Jun 2021

%\cite{1611.03874}
\bibitem{1611.03874}
M.~G.~Aartsen \textit{et al.} [IceCube],
``The contribution of Fermi-2LAC blazars to the diffuse TeV-PeV neutrino flux,''
Astrophys. J. \textbf{835}, no.1, 45 (2017)
%doi:10.3847/1538-4357/835/1/45
[arXiv:1611.03874 [astro-ph.HE]].
%173 citations counted in INSPIRE as of 01 Jul 2021

%\cite{1810.02823}
\bibitem{1810.02823}
D.~Hooper, T.~Linden and A.~Vieregg,
``Active Galactic Nuclei and the Origin of IceCube's Diffuse Neutrino Flux,''
JCAP \textbf{02}, 012 (2019)
%doi:10.1088/1475-7516/2019/02/012
[arXiv:1810.02823 [astro-ph.HE]].
%26 citations counted in INSPIRE as of 01 Jul 2021

%\cite{1904.06371}
\bibitem{1904.06371}
C.~Yuan, K.~Murase and P.~M\'esz\'aros,
``Complementarity of Stacking and Multiplet Constraints on the Blazar Contribution to the Cumulative High-Energy Neutrino Intensity,''
Astrophys. J. \textbf{890}, 25 (2020)
%doi:10.3847/1538-4357/ab65ea
[arXiv:1904.06371 [astro-ph.HE]].
%14 citations counted in INSPIRE as of 05 Jul 2021

%\cite{2001.00930}
\bibitem{2001.00930}
A.~Plavin, Y.~Y.~Kovalev, Y.~A.~Kovalev and S.~Troitsky,
``Observational Evidence for the Origin of High-energy Neutrinos in Parsec-scale Nuclei of Radio-bright Active Galaxies,''
Astrophys. J. \textbf{894}, no.2, 101 (2020)
%doi:10.3847/1538-4357/ab86bd
[arXiv:2001.00930 [astro-ph.HE]].
%16 citations counted in INSPIRE as of 26 Jun 2021

%\cite{2009.08914}
\bibitem{2009.08914}
A.~V.~Plavin, Y.~Y.~Kovalev, Y.~A.~Kovalev and S.~V.~Troitsky,
``Directional Association of TeV to PeV Astrophysical Neutrinos with Radio Blazars,''
Astrophys. J. \textbf{908}, no.2, 157 (2021)
%doi:10.3847/1538-4357/abceb8
[arXiv:2009.08914 [astro-ph.HE]].
%9 citations counted in INSPIRE as of 26 Jun 2021

%\cite{2103.12813}
\bibitem{2103.12813}
B.~Zhou, M.~Kamionkowski and Y.~f.~Liang,
%``Search for High-Energy Neutrino Emission from Radio-Bright AGN,''
Phys. Rev. D \textbf{103}, no.12, 123018 (2021)
doi:10.1103/PhysRevD.103.123018
[arXiv:2103.12813 [astro-ph.HE]].
%0 citations counted in INSPIRE as of 06 Jul 2021

%\cite{1706.02175}
\bibitem{1706.02175}
N.~Senno, K.~Murase and P.~M\'esz\'aros,
``Constraining high-energy neutrino emission from choked jets in stripped-envelope supernovae,''
JCAP \textbf{01}, 025 (2018)
%doi:10.1088/1475-7516/2018/01/025
[arXiv:1706.02175 [astro-ph.HE]].
%14 citations counted in INSPIRE as of 26 Jun 2021

%\cite{1809.09610}
\bibitem{1809.09610}
A.~Esmaili and K.~Murase,
``Constraining high-energy neutrinos from choked-jet supernovae with IceCube high-energy starting events,''
JCAP \textbf{12}, 008 (2018)
%doi:10.1088/1475-7516/2018/12/008
[arXiv:1809.09610 [hep-ph]].
%12 citations counted in INSPIRE as of 26 Jun 2021

%\cite{2003.12071}
\bibitem{2003.12071}
M.~G.~Aartsen \textit{et al.} [IceCube],
``IceCube Search for High-Energy Neutrino Emission from TeV Pulsar Wind Nebulae,''
Astrophys. J. \textbf{898}, no.2, 117 (2020)
%doi:10.3847/1538-4357/ab9fa0
[arXiv:2003.12071 [astro-ph.HE]].
%10 citations counted in INSPIRE as of 26 Jun 2021

%\cite{2008.04323}
\bibitem{2008.04323}
M.~G.~Aartsen \textit{et al.} [IceCube-Gen2],
``IceCube-Gen2: the window to the extreme Universe,''
J. Phys. G \textbf{48}, no.6, 060501 (2021)
%doi:10.1088/1361-6471/abbd48
[arXiv:2008.04323 [astro-ph.HE]].
%47 citations counted in INSPIRE as of 02 Jul 2021

%\cite{1108.2509}
\bibitem{1108.2509}
V.~Berezinsky, E.~Sabancilar and A.~Vilenkin,
``Extremely High Energy Neutrinos from Cosmic Strings,''
Phys. Rev. D \textbf{84}, 085006 (2011)
%doi:10.1103/PhysRevD.84.085006
[arXiv:1108.2509 [astro-ph.CO]].
%46 citations counted in INSPIRE as of 29 Jun 2021

%\cite{1206.2924}
\bibitem{1206.2924}
C.~Lunardini and E.~Sabancilar,
``Cosmic Strings as Emitters of Extremely High Energy Neutrinos,''
Phys. Rev. D \textbf{86}, 085008 (2012)
%doi:10.1103/PhysRevD.86.085008
[arXiv:1206.2924 [astro-ph.CO]].
%29 citations counted in INSPIRE as of 26 Jun 2021

%\cite{1312.4573}
\bibitem{1312.4573}
J.~M.~Hyde, A.~J.~Long and T.~Vachaspati,
``Dark Strings and their Couplings to the Standard Model,''
Phys. Rev. D \textbf{89}, 065031 (2014)
%doi:10.1103/PhysRevD.89.065031
[arXiv:1312.4573 [hep-ph]].
%28 citations counted in INSPIRE as of 26 Jun 2021

%\cite{1405.7679}
\bibitem{1405.7679}
A.~J.~Long, J.~M.~Hyde and T.~Vachaspati,
``Cosmic Strings in Hidden Sectors: 1. Radiation of Standard Model Particles,''
JCAP \textbf{09}, 030 (2014)
%doi:10.1088/1475-7516/2014/09/030
[arXiv:1405.7679 [hep-ph]].
%25 citations counted in INSPIRE as of 26 Jun 2021

%\cite{1409.6979}
\bibitem{1409.6979}
A.~J.~Long and T.~Vachaspati,
``Cosmic Strings in Hidden Sectors: 2. Cosmological and Astrophysical Signatures,''
JCAP \textbf{12}, 040 (2014)
%doi:10.1088/1475-7516/2014/12/040
[arXiv:1409.6979 [hep-ph]].
%21 citations counted in INSPIRE as of 26 Jun 2021

%\cite{Vachaspati:1984gt}
\bibitem{Vachaspati:1984gt}
T.~Vachaspati and A.~Vilenkin,
``Gravitational Radiation from Cosmic Strings,''
Phys. Rev. D \textbf{31}, 3052 (1985)
%doi:10.1103/PhysRevD.31.3052
%411 citations counted in INSPIRE as of 26 Jun 2021

%\cite{Hindmarsh:1990xi}
\bibitem{Hindmarsh:1990xi}
M.~Hindmarsh,
``Gravitational radiation from kinky infinite strings,''
Phys. Lett. B \textbf{251}, 28-33 (1990)
%doi:10.1016/0370-2693(90)90226-V
%51 citations counted in INSPIRE as of 26 Jun 2021

%\cite{Allen:1991bk}
\bibitem{Allen:1991bk}
B.~Allen and E.~P.~S.~Shellard,
``Gravitational radiation from cosmic strings,''
Phys. Rev. D \textbf{45}, 1898-1912 (1992)
%doi:10.1103/PhysRevD.45.1898
%66 citations counted in INSPIRE as of 26 Jun 2021

%\cite{0911.2655}
\bibitem{0911.2655}
T.~Vachaspati,
``Cosmic Rays from Cosmic Strings with Condensates,''
Phys. Rev. D \textbf{81}, 043531 (2010)
%doi:10.1103/PhysRevD.81.043531
[arXiv:0911.2655 [astro-ph.CO]].
%36 citations counted in INSPIRE as of 26 Jun 2021

%\cite{gr-qc/0104026}
\bibitem{gr-qc/0104026}
T.~Damour and A.~Vilenkin,
``Gravitational wave bursts from cusps and kinks on cosmic strings,''
Phys. Rev. D \textbf{64}, 064008 (2001)
%doi:10.1103/PhysRevD.64.064008
[arXiv:gr-qc/0104026 [gr-qc]].
%278 citations counted in INSPIRE as of 01 Jul 2021

%\cite{1911.12066}
\bibitem{1911.12066}
P.~Auclair, D.~A.~Steer and T.~Vachaspati,
``Particle emission and gravitational radiation from cosmic strings: observational constraints,''
Phys. Rev. D \textbf{101}, no.8, 083511 (2020)
%doi:10.1103/PhysRevD.101.083511
[arXiv:1911.12066 [hep-ph]].
%15 citations counted in INSPIRE as of 05 Jul 2021

%\cite{2005.05332}
\bibitem{2005.05332}
C.~Creque-Sarbinowski, J.~Hyde and M.~Kamionkowski,
``Resonant neutrino self-interactions,''
Phys. Rev. D \textbf{103}, no.2, 023527 (2021)
%doi:10.1103/PhysRevD.103.023527
[arXiv:2005.05332 [hep-ph]].
%2 citations counted in INSPIRE as of 26 Jun 2021

%\cite{Ema:2013nda}
\bibitem{1312.3501}
Y.~Ema, R.~Jinno and T.~Moroi,
``Cosmic-Ray Neutrinos from the Decay of Long-Lived Particle and the Recent IceCube Result,''
Phys. Lett. B \textbf{733}, 120-125 (2014)
%doi:10.1016/j.physletb.2014.04.021
[arXiv:1312.3501 [hep-ph]].
%73 citations counted in INSPIRE as of 17 May 2022


%\cite{0911.0682}
\bibitem{0911.0682}
K.~Jones-Smith, H.~Mathur and T.~Vachaspati,
``Aharonov-Bohm Radiation,''
Phys. Rev. D \textbf{81}, 043503 (2010)
%doi:10.1103/PhysRevD.81.043503
[arXiv:0911.0682 [hep-th]].
%18 citations counted in INSPIRE as of 26 Jun 2021

%\cite{Alford:1988sj}
\bibitem{Alford:1988sj}
M.~G.~Alford and F.~Wilczek,
``Aharonov-Bohm Interaction of Cosmic Strings with Matter,''
Phys. Rev. Lett. \textbf{62}, 1071 (1989)
%doi:10.1103/PhysRevLett.62.1071
%218 citations counted in INSPIRE as of 26 Jun 2021

%\cite{Berezinsky:2000up}
\bibitem{hep-ph/0009053}
V.~Berezinsky and M.~Kachelriess,
``Monte Carlo simulation for jet fragmentation in SUSY QCD,''
%Phys. Rev. D \textbf{63}, 034007 (2001)
%doi:10.1103/PhysRevD.63.034007
[arXiv:hep-ph/0009053 [hep-ph]].
%98 citations counted in INSPIRE as of 17 May 2022

%\cite{Sarkar:2001se}
\bibitem{hep-ph/0108098}
S.~Sarkar and R.~Toldra,
``The High-energy cosmic ray spectrum from relic particle decay,''
Nucl. Phys. B \textbf{621}, 495-520 (2002)
%doi:10.1016/S0550-3213(01)00565-X
[arXiv:hep-ph/0108098 [hep-ph]].
%168 citations counted in INSPIRE as of 17 May 2022



%\cite{Barbot:2002gt}
\bibitem{hep-ph/0211406}
C.~Barbot and M.~Drees,
``Detailed analysis of the decay spectrum of a super heavy X particle,''
Astropart. Phys. \textbf{20}, 5-44 (2003)
%doi:10.1016/S0927-6505(03)00134-8
[arXiv:hep-ph/0211406 [hep-ph]].
%79 citations counted in INSPIRE as of 17 May 2022

%\cite{Aloisio:2003xj}
\bibitem{hep-ph/0307279}
R.~Aloisio, V.~Berezinsky and M.~Kachelriess,
``Fragmentation functions in SUSY QCD and UHECR spectra produced in top - down models,''
Phys. Rev. D \textbf{69}, 094023 (2004)
%doi:10.1103/PhysRevD.69.094023
[arXiv:hep-ph/0307279 [hep-ph]].
%105 citations counted in INSPIRE as of 17 May 2022


%\cite{hep-ph/9803414}
\bibitem{hep-ph/9803414}
E.~J.~Copeland, T.~W.~B.~Kibble and D.~A.~Steer,
``The Evolution of a network of cosmic string loops,''
Phys. Rev. D \textbf{58}, 043508 (1998)
%doi:10.1103/PhysRevD.58.043508
[arXiv:hep-ph/9803414 [hep-ph]].
%24 citations counted in INSPIRE as of 30 Jun 2021

%\cite{gr-qc/0001023}
\bibitem{gr-qc/0001023}
T.~Damour, B.~R.~Iyer and B.~S.~Sathyaprakash,
``Frequency domain P approximant filters for time truncated inspiral gravitational wave signals from compact binaries,''
Phys. Rev. D \textbf{62}, 084036 (2000)
%doi:10.1103/PhysRevD.62.084036
[arXiv:gr-qc/0001023 [gr-qc]].
%132 citations counted in INSPIRE as of 29 Jun 2021

%\cite{Planck:2018vyg}
\bibitem{1807.06209}
N.~Aghanim \textit{et al.} [Planck],
``Planck 2018 results. VI. Cosmological parameters,''
Astron. Astrophys. \textbf{641}, A6 (2020)
[erratum: Astron. Astrophys. \textbf{652}, C4 (2021)]
%doi:10.1051/0004-6361/201833910
[arXiv:1807.06209 [astro-ph.CO]].
%7529 citations counted in INSPIRE as of 22 Apr 2022
%\cite{IceCube:2013low}
\bibitem{1311.5238}
M.~G.~Aartsen \textit{et al.} [IceCube],
``Evidence for High-Energy Extraterrestrial Neutrinos at the IceCube Detector,''
Science \textbf{342}, 1242856 (2013)
%doi:10.1126/science.1242856
[arXiv:1311.5238 [astro-ph.HE]].
%1382 citations counted in INSPIRE as of 24 May 2022

%\cite{Yamaguchi:2005gp}
\bibitem{hep-ph/0503227}
M.~Yamaguchi,
``Cosmological evolution of cosmic strings with time dependent tension,''
Phys. Rev. D \textbf{72}, 043533 (2005)
%doi:10.1103/PhysRevD.72.043533
[arXiv:hep-ph/0503227 [hep-ph]].
%12 citations counted in INSPIRE as of 29 Apr 2022


%\cite{ANITA:2019wyx}
\bibitem{2012.07945}
P.~W.~Gorham \textit{et al.} [ANITA],
``Constraints on the ultrahigh-energy cosmic neutrino flux from the fourth flight of ANITA,''
Phys. Rev. D \textbf{99}, no.12, 122001 (2019)
%doi:10.1103/PhysRevD.99.122001
[arXiv:1902.04005 [astro-ph.HE]].
%69 citations counted in INSPIRE as of 17 May 2022

%\cite{POEMMA:2020ykm}
\bibitem{1902.04005}
A.~V.~Olinto \textit{et al.} [POEMMA],
``The POEMMA (Probe of Extreme Multi-Messenger Astrophysics) observatory,''
JCAP \textbf{06}, 007 (2021)
%doi:10.1088/1475-7516/2021/06/007
[arXiv:2012.07945 [astro-ph.IM]].
%46 citations counted in INSPIRE as of 17 May 2022

%\cite{IceCube-Gen2:2020qha}
\bibitem{2008.04323}
M.~G.~Aartsen \textit{et al.} [IceCube-Gen2],
``IceCube-Gen2: the window to the extreme Universe,''
J. Phys. G \textbf{48}, no.6, 060501 (2021)
%doi:10.1088/1361-6471/abbd48
[arXiv:2008.04323 [astro-ph.HE]].
%172 citations counted in INSPIRE as of 19 May 2022

\end{thebibliography}
\end{document}